\documentclass{article} 

\usepackage{nips,times}
\usepackage{hyperref}
\usepackage{url}
\usepackage[sort,comma,authoryear]{natbib}
\usepackage{bbm}
\usepackage{amsmath}
\usepackage{amssymb}
\usepackage{graphicx}
\usepackage{booktabs}
\usepackage{caption}

\newcommand{\mvec}[1]{\ensuremath{\mathbf{#1}}}
\newcommand{\msvec}[1]{\ensuremath{\boldsymbol{#1}}}

\newcommand{\mcal}[1]{\ensuremath{\mathcal{#1}}}

\newcommand{\myfrac}[3][0pt]{\genfrac{}{}{}{}{\raisebox{#1}{$#2$}}{\raisebox{-#1}{$#3$}}}

\title{Higher order methylation features for clustering and prediction in epigenomic studies}

\author{
Chantriolnt-Andreas~Kapourani\\
IANC, School of Informatics\\
University of Edinburgh\\
Edinburgh, EH8 9AB, UK \\
\texttt{C.A.Kapourani@ed.ac.uk} \\
\And
Guido~Sanguinetti\thanks{To whom correspondence should be addressed.} \\
IANC, School of Informatics \\
University of Edinburgh\\
Edinburgh, EH8 9AB, UK \\
\texttt{G.Sanguinetti@ed.ac.uk} \\
}

\nipsfinalcopy 

\begin{document}

\maketitle

\begin{abstract}
\textbf{Motivation:} DNA methylation is an intensely studied epigenetic mark, yet its functional role is incompletely understood. Attempts to quantitatively associate average DNA methylation to gene expression yield poor correlations outside of the well-understood methylation-switch at CpG islands.\\
\textbf{Results:} Here we use probabilistic machine learning to extract higher order features associated with the methylation profile across a defined region. These features quantitate precisely notions of shape of a methylation profile, capturing spatial correlations in DNA methylation across genomic regions. Using these higher order features across promoter-proximal regions, we are able to construct a powerful machine learning predictor of gene expression, significantly improving upon the predictive power of average DNA methylation levels. Furthermore, we can use higher order features to cluster promoter-proximal regions, showing that five major patterns of methylation occur at promoters across different cell lines, and we provide evidence that methylation beyond CpG islands may be related to regulation of gene expression. Our results support previous reports of a functional role of spatial correlations in methylation patterns, and provide a mean to quantitate such features for downstream analyses.\\
\textbf{Availability:} \href{https://github.com/andreaskapou/BPRMeth}{https://github.com/andreaskapou/BPRMeth} \\
\end{abstract}

\section{Introduction}
DNA methylation is a well studied, heritable epigenetic modification which plays an important role in gene regulatory mechanisms. It is associated with a broad range of biological processes of direct clinical relevance, including X-chromosome inactivation, genomic imprinting, silencing of repetitive DNA and carcinogenesis \citep{Feinberg1983, Li1993, Baylin2011}. Methylation occurs when a methyl group is attached to a DNA nucleotide. In vertebrate genomes, methylation is observed almost exclusively on 5-methylcytosine (5-mC) residues in the context of CpG dinucleotides. Due to increased vulnerability of 5-mC to randomly deaminate into thymine, most of the genome is depleted from CpG dinucleotides, except from small CpG-rich regions, termed CpG islands (CGIs) \citep{Bird2002}. Hyper-methylation of CGIs near promoter regions is generally associated with transcriptional repression; however, outside of this well documented case, the association between DNA methylation across promoter-proximal regions and transcript abundance is considerably weaker and poorly understood \citep{Jones2012}.

Recent advances in high-throughput sequencing technology have made it possible to measure the methylation level of cytosines on a genome-wide scale with single nucleotide resolution. Sodium bisulphite treatment of DNA followed by sequencing (BS-seq) efficiently converts unmethylated cytosines to uracils (which are subsequently amplified as thymines by PCR) and leaves the 5-mCs unmodified \citep{Krueger2012}. To obtain DNA methylation levels, reads are aligned to a reference genome allowing changes of cytosines to thymines during the mapping procedure. A variant of BS-seq technology, termed Reduced Representation Bisulphite Sequencing (RRBS) \citep{Meissner2005}, uses methylation-sensitive restriction enzymes to cleave the DNA at specific loci before bisulphite treatment. This results in measuring in greater coverage and at lower cost the methylation level of CpG-rich regions genome-wide.

Despite the widespread take up of BS-seq technology, statistical modelling of such data is still challenging, yet it is crucial in order to uncover biological regulatory mechanisms. Analysis of BS-seq data has mainly focused on identifying differentially methylated regions (DMRs) across different conditions. Some notable DMR methods are BSmooth \citep{Hansen2012}, Bi-Seq \citep{Hebestreit2013} and $M^{3}D$ \citep{Mayo2015}. While DMR detection methods are often crucial ingredients in exploratory data analysis pipelines, they do not provide a clear platform to quantitatively understand the relationship between DNA methylation and gene expression. Most studies use DMR detection as a prefilter, and then simply correlate mean methylation levels across each region (often taken to be promoter-proximal) with gene expression. Adopting this simple approach, genome-wide studies \citep{Hansen2011, Bock2012a} have reported only modest correlation between average DNA-methylation and gene expression {(Pearson's correlation coefficient $r \approx$ -0.3)}.

\begin{figure}[!pb]
  \begin{center}
  \begin{minipage}{0.39\textwidth}
    \includegraphics[width=\textwidth]{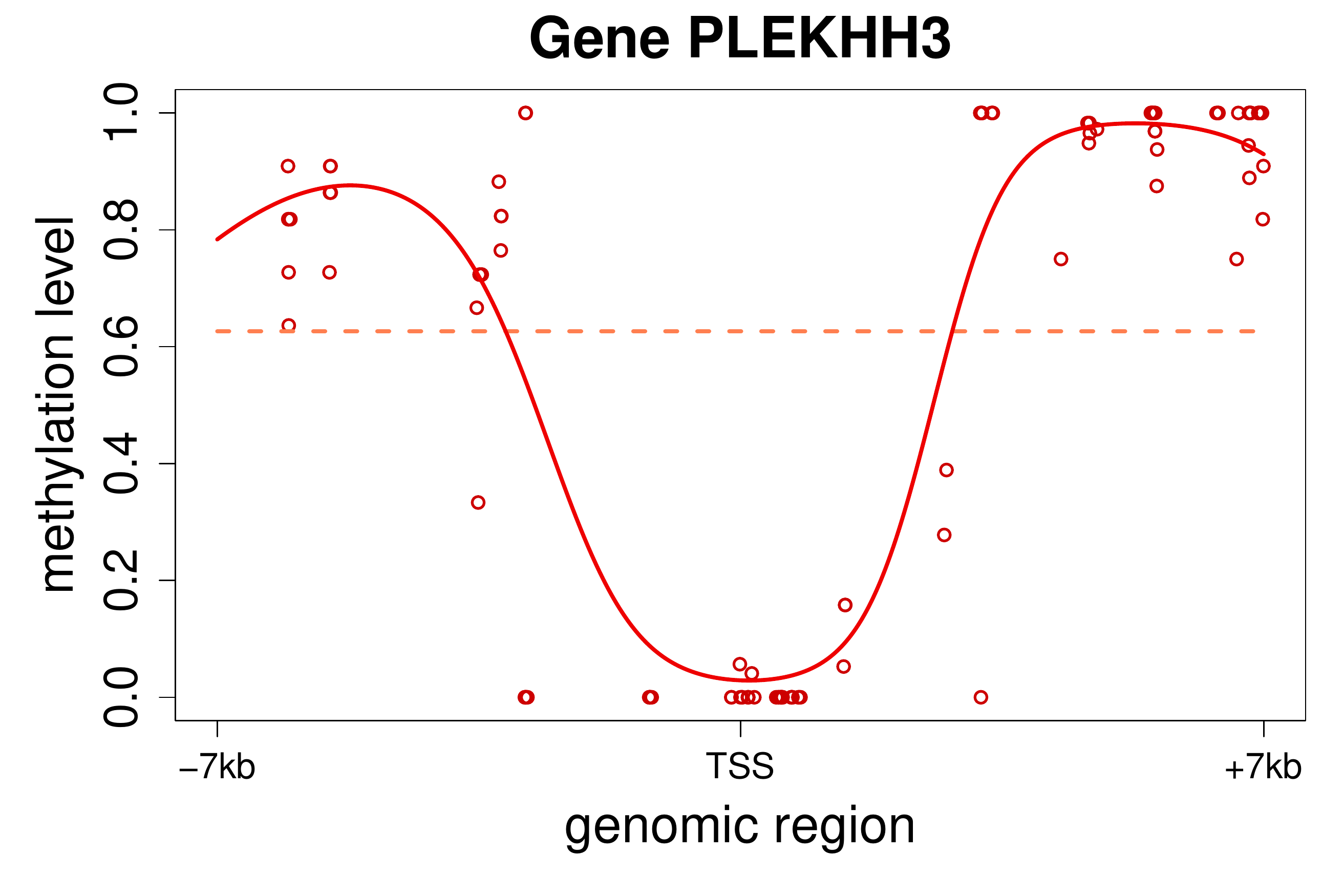}
  \end{minipage}%
  \hspace*{0.8pt}
  \begin{minipage}{0.39\textwidth}
    \includegraphics[width=\textwidth]{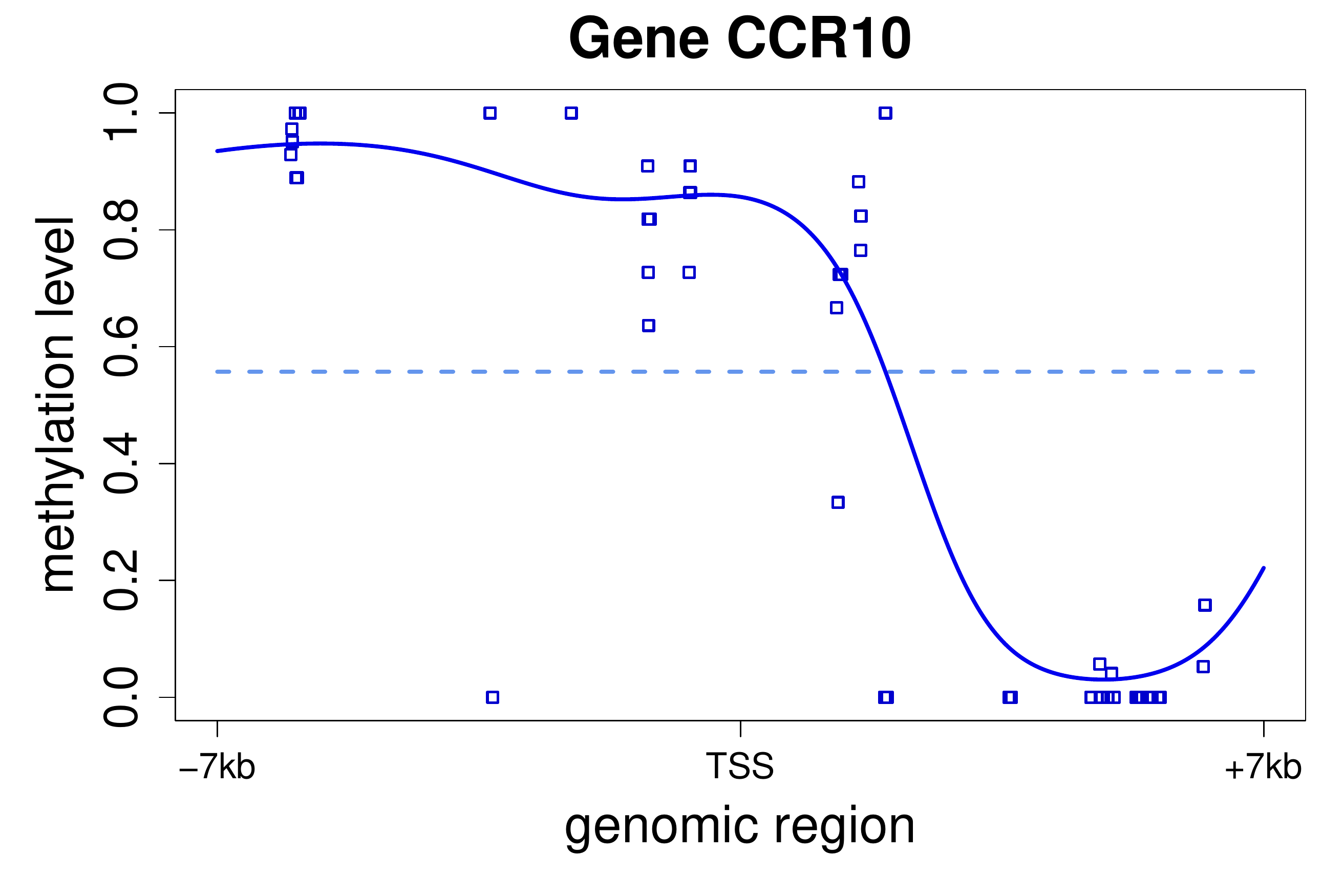}
  \end{minipage}
  \caption{\small{Methylation patterns for the PLEKHH3 and CCR10 genes from the K562 cell line over $\pm7kb$ promoter region. Each point represents the relative CpG location w.r.t. TSS and the corresponding DNA methylation level. The dashed horizontal lines show the average methylation level. The shapes of methylation profiles are very different, however the average methylation level cannot explain them. Also, note that there are no CpG measurements in the (-$6kb$, -$4kb$) region for the CCR10 gene, and the learned methylation profiles can be thought as imputing the missing values by taking into consideration the spatial co-dependence of nearby CpGs.}}
  \label{fig:methylation-profiles}
  \end{center}
\end{figure}

In this paper, we argue that part of the difficulty in quantitatively associating methylation levels with gene expression resides in the simplistic encoding of DNA methylation  across a region as a simple average. DNA methylation often displays reproducible, spatially correlated patterns ({\it profiles}); Figure~\ref{fig:methylation-profiles} shows two examples from an ENCODE data sets \citep{Dunham2012}. This spatial reproducibility was exploited by \cite{Mayo2015} to provide more powerful tests for DMR, and by \cite{Vanderkraats2013} to group genes with similar differential methylation patterns and corresponding expression changes. These results suggest that a precise quantification of the spatial variability in the DNA methylation mark may aid the quest to quantitatively  understand the interplay between methylation and transcription. We propose a probabilistic model of methylation profiles, based on latent variable models, which allows us to associate with each region of interest a set of features capturing precisely the methylation profile across the region. We then show that, using such features, we can construct an accurate machine learning predictor of gene expression from DNA methylation, achieving test correlations twice as large as previously reported.

The rest of the paper is organised as follows: we start off by providing a high-level description of our approach. We then describe precisely the statistical methodology we propose. We illustrate our approach on three ENCODE data sets, showing that higher order features allow much more accurate predictions of gene expression. We also show how such features can be used to cluster regions according to their methylation profiles, and show that five prototypical methylation profiles appear to explain most variability in promoter-proximal methylation in human cell lines.

\section{Approach}
In this paper, we propose a novel probabilistic machine learning methodology to quantify the profile of DNA methylation across genomic regions from BS-seq data. Our motivation is practical: inspection of many BS-seq data sets reveals that methylation levels across promoter-proximal regions often show reproducible, spatially correlated profiles. Figure~\ref{fig:methylation-profiles} shows two example promoter-proximal regions which clearly display such spatial correlations, resulting in characteristic methylation shapes. We propose a method to quantitate such qualitative information.

The method is based on a Generalised Linear Model of basis function regression coupled with a Binomial observation likelihood, and allows us to associate each region with a set of basis function coefficients which capture the methylation profile. We show how such higher order features can then be used in downstream analysis to yield a significantly improved estimate of the correlation between methylation and gene expression, and to identify prototypical methylation profiles across promoter regions.

\section{Methods}

\subsection{Modelling DNA methylation profiles}
As in all HTS-based assays, the output of a BS-seq experiment is a set of reads aligned to the genome; the main difference is that the bisulphite treatment changes to thymine any unmethylated cytosine. Thus, the same base on the genome will appear as cytosine on some reads, and as thymine on others; the ratio of reads containing a cytosine readout to total reads gives a measurement of the sample methylation level. This measurement process at a single cytosine can be naturally modelled with a Binomial distribution, where the number of successes represents the number of reads on which the cytosine actually appears as C, and the number of attempts is the total number of reads mapping to the specific site. Let \emph{t} be the total number of reads that are mapped to a specific CpG site, and let \emph{m} of these reads to contain methylated cytosines. Then, for each CpG site we assume that $m \sim \mcal{B}inom(t,p)$, where $p$ is the unknown methylation level. 

In this paper, and in many practical studies, we are interested in learning the methylation patterns of fixed-width genomic regions, e.g. promoters. Hence, each genomic region $i (i = 1, \ldots, N)$ can be represented as a vector of CpG locations $\mvec{x}_{i}$, where each entry corresponds to the location of the CpG in the genomic region, relative to a reference point such as the Transcription Start Site (TSS). It should be noted that the vector lengths $L_{i}$ may vary between different genomic regions, since they depend on the number of actual CpG dinucleotides found in each region. For each region $i$, we also have a vector of observations $\mvec{y}_{i}$, containing the methylation levels of the corresponding CpG sites; each entry consists of the tuple $y_{il} = (m_{il}, t_{il})$, where, $m_{il}$ is the number of 5-mC reads mapped to the $l$-th CpG site in region $i$, and $t_{il}$ corresponds to the total number of reads.

Direct comparison of the observation vectors $\mvec{y}_{i}$ for different regions is complicated due to the variability in the vector lengths. To enable comparisons between these regions, we formulate our problem as a regression problem, where the methylation profile of each genomic region is modelled as a linear combination of a set of latent basis functions. Let $f(\mvec{x}_{i})$ be a latent function representing the methylation profile for genomic region $i$. Since the observed methylation data contain the proportion of methylated reads out of the total reads for each CpG site, each entry of the vector $\mvec{y}_{i}$ takes values in the $[0, 1]$ interval. Thus, we introduce an unconstrained latent function $g(\mvec{x}_{i})$ defined so that $f(\mvec{x}_{i})$ is the probit transformation of $g(\mvec{x}_{i})$: $f(\mvec{x}_{i}) = \Phi \big(g(\mvec{x}_{i})\big)$, where $\Phi(\cdot)$ denotes the cumulative distribution function (cdf) of the standard normal distribution. Let $\mvec{f}_{i} = f(\mvec{x}_{i})$ and $\mvec{g}_{i} = g(\mvec{x}_{i})$ be shorthand for the values of the latent functions. 

Given the values of the latent function $\mvec{f}_{i}$ for region $i$, the observations $y_{il}$ for each CpG site are independent and identically distributed Binomial variables, so we can define the joint log-likelihood for region $i$ in factorised form:
\begin{equation}\label{eq:bpr-likelihood}
\log p(\mvec{y}_{i} | \mvec{f}_{i}) = \sum\limits_{l = 1}^{L_{i}} \log \bigg\lbrace \mcal{B}inom\big(m_{il} | t_{il}, \Phi(g_{il})\big) \bigg\rbrace
\end{equation}

From its final form, we refer to this observation model as the Binomial distributed Probit Regression (BPR) likelihood function. Notice that the BPR model explicitly accounts for the coverage variability across CpG sites through the use of the Binomial observation model: as the variance of a binomial distribution decreases rapidly with the number of attempts, the model will be very strongly constrained by highly covered sites. Hence, it handles in a principled way the uncertainty present in low coverage reads during the analysis of BS-seq data.

\subsection{Feature Extraction}
To constrain the latent function $\mvec{g}_{i}$ we assume it is given as a linear combination of fixed non-linear basis functions $h_{j}(\cdot)$ of the input space $\mvec{x}_{i}$, of the form:
\begin{equation}\label{eq:linear-comb-basis-func}
\mvec{g}_{i}(\mvec{x}_{i}, \mvec{w}_{i}) = \sum\limits_{j = 0}^{M-1} w_{j} h_{j}(\mvec{x}_{i}) =  \mvec{H}_{i} \mvec{w}_{i}
\end{equation}
where $\mvec{w}_{i} = (w_{i,0}, \ldots, w_{i,M-1})^T$, $\mvec{H}_{i}$ is the $L_{i} \times M$ design matrix, whose elements are given by $\mvec{H}_{ilj} = h_{j}(x_{il})$, and $M$ denotes the total number of basis functions. Hence, its probit transformation $\mvec{f}_{i}$ is given by:
\begin{equation}
\mvec{f}_{i}(\mvec{x}_{i}, \mvec{w}_{i}) = \Phi\big(\mvec{g}_{i}(\mvec{x}_{i}, \mvec{w}_{i})) = \Phi\big(\mvec{H}_{i} \mvec{w}_{i}\big)
\end{equation}

One should note that even though the function $\mvec{g}_{i}$ is linear with respect to the parameters $\mvec{w}_{i}$, the latent function $\mvec{f}_{i}$ is non linear due to the presence of the probit transformation. In this study, we consider Radial Basis Functions (RBFs); for a single input variable $x$, the RBF takes the form $h_{j}(x) = exp(-\gamma || x - \mu_{j} ||^2)$, where $\mu_{j}$ denotes the location of the $j^{th}$ basis function in the input space and $\gamma$ controls the spatial scale. 

Learning the methylation profiles $\mvec{f}_{i}$ for each genomic region, is equivalent to optimising the model parameters $\mvec{w}_{i}$. The parameters $\mvec{w}_{i}$ can be considered as the extracted features which quantitate precisely notions of shape of a methylation profile. Optimising $\mvec{w}_{i}$ involves maximising Equation~(\ref{eq:bpr-likelihood}) for each genomic region; however, by increasing the number of basis functions, we also increase the resolution for the shape of the methylation profiles, which might lead to overfitting. To ameliorate this issue, we maximise a penalised version of Equation~(\ref{eq:bpr-likelihood}), by adding a regularisation term $\mcal{E}(\mvec{w}_{i})$ to the log-likelihood function which will encourage the weights to decay to zero:
\vspace*{3pt}
\begin{equation}\label{eq:bpr-lik-penalized}
J(\mvec{w}_{i}) = \log p(\mvec{y}_{i} | \mvec{f}_{i}, \mvec{w}_{i}) - \mcal{E}(\mvec{w}_{i})
\end{equation}
where $\mcal{E}(\mvec{w}_{i}) = \frac{1}{2} \mvec{w}^{T}_{i}\mvec{w}_{i}$ is the squared two-norm.
This approach is known as ridge regression or weight decay.

Direct maximisation of $J(\mvec{w}_{i})$ w.r.t parameters $\mvec{w}_{i}$ is intractable due to presence of the probit transformation. We use gradient-based numerical optimisation techniques, such as Conjugate Gradient (CG) and BFGS, to perform the optimisation.

\subsection{Predicting gene expression}
The extracted higher-order methylation features across promoter-proximal regions can be used for downstream analysis, such as predicting transcript abundance, or performing clustering in order to learn prototypical methylation patterns that occur at promoters across different cell lines.

To quantitatively predict expression at each promoter region, we construct a regression model by taking as input the higher-order methylation features extracted from each promoter-proximal region. The performance of the regression model is evaluated by computing the root-mean squared error (RMSE) and the Pearson's correlation coefficient (r) between the predicted and the measured (log-transformed) gene expression levels. We compare our proposed model's performance with the standard approach \citep{Hansen2011, Bock2012a} which uses the average methylation level across a region as input feature (this approach can be thought of as fitting a constant function across each genomic region). We have tested both a linear regression model and a variety of non-linear models, such as SVM regression, Random Forests and Multivariate Adaptive Regression Splines (MARS) \citep{Friedman1991}. The SVM regression is consistently better than the other regression models, hence, we choose this model for the rest of our analysis.

In addition to the methylation profile features, we consider two supplementary sources of information which could plausibly act as confounders in the predictions. The first feature accounts for the goodness of fit of each methylation profile to the observed methylation data using the RMSE as error measure, intuitively quantitating the noisiness in the methylation profile. The second feature considers the number of CpG dinucleotides present in each promoter region. It is thought that CpG density may play a functional role in controlling gene expression, with the main evidence being the existence of CpG islands \citep{Deaton2011}.

\subsection{Clustering methylation profiles}
To cluster methylation profiles we consider a mixture modelling approach \citep{McLachlan2004}. We assume that the methylation profiles $\mvec{f}$ can be partitioned into at most K clusters, and each cluster $k$ can be modelled separately using the BPR likelihood as our observation model. The log-likelihood for the mixture model is defined as:
\begin{equation}\label{eq:mixture-model}
p(\mvec{y} | \msvec{\Theta}) = \sum\limits_{i = 1}^{N} \log \bigg\lbrace\sum_{k = 1}^{K} \pi_{k} p(\mvec{y}_{i} | \mvec{f}_{i}, \mvec{w}_{k}, z_{i} = k)\bigg\rbrace 
\end{equation}
where $\msvec{\Theta} = (\pi_{1}, \ldots , \pi_{k}, \mvec{w}_{1}, \ldots, \mvec{w}_{k})$, $\pi_{k}$ are the mixing proportions (with $\pi_{k} \in (0,1) \; \forall k$ and $\sum_{k}\pi_{k} = 1$), $\mvec{w}_{k}$ are the methylation profile parameters and $z_{i}$ are the latent variables denoting to which cluster each genomic region belongs. To avoid cluttering the notation, we will omit the dependence of the observation model on the latent variables $z_{i}$.

\subsubsection*{Parameter Estimation}
To estimate the model parameters $\msvec{\Theta} = (\pi_{1}, \ldots , \pi_{k}, \mvec{w}_{1}, \ldots, \mvec{w}_{k})$, the Expectation Maximization (EM) algorithm \citep{Dempster1977} is considered. EM is a general iterative algorithm for computing maximum likelihood estimates when there are missing or latent variables, as in the case of mixture models. EM alternates between inferring the latent variables given the parameters (E-step), and optimising the parameters given the posterior statistics of the latent variables (M-step). Formally, during the E-step we compute the responsibility that component $k$ takes for explaining observations $\mvec{y}_{i}$:
\begin{equation} \label{eq:compute-resp}
 \gamma(z_{ik}) = \myfrac[4pt]{\pi_{k} p(\mvec{y}_{i} | \mvec{f}_{i}, \mvec{w}_{k})}{\sum_{j = 1}^{K} \pi_{j} p(\mvec{y}_{i} | \mvec{f}_{i}, \mvec{w}_{j})}
\end{equation}
The M-step consists of updating the model parameters so as to maximise the expected complete data log-likelihood. The mixing proportions $\pi_{k}$ are updated as follows:
\begin{equation} \label{eq:update-mix-prop}
\pi_{k} = \frac{1}{N}\sum\limits_{i = 1}^{N} \gamma(z_{ik})
\end{equation}
To re-estimate the observation model parameters $\mvec{w}_{k}$ we need to optimise the following quantity:
\begin{equation} \label{eq-em-opt}
\ell(\mvec{w}_{k}) = \sum_{i}  \gamma(z_{ik}) \sum_{l}  \log \bigg\lbrace \mathcal{B}inom \big(m_{il} | t_{il}, \Phi(g_{il}; \mvec{w}_{k})\big) \bigg \rbrace
\end{equation}
However, direct optimisation of $\ell(\mvec{w}_{k})$ w.r.t parameters $\mvec{w}_{k}$ is intractable, thus, we resort again to numerical optimisation strategies.  This variant of EM algorithm is known as Generalised EM, or GEM, and it is proved to converge to the maximum likelihood estimate \citep{Wu1983}. It should be noted that the penalised version of the BPR likelihood, given in Equation (\ref{eq:bpr-lik-penalized}), can be easily incorporated in the clustering approach.

\section{Data Sets}
To evaluate the performance of the proposed methodology we use real datasets that are publicly available from the ENCODE project consortium \citep{Dunham2012}. More specifically, the following three Tier 1 cell lines are used:
\begin{enumerate}
\item{K562 immortalized cell line, coming from a human female with chronic myelogenous leukemia.}
\item{GM12878 lymphoblastoid cell line, produced from the blood of a female donor with northern and western European ancestry by EBV transformation.}
\item{H1-hESC embryonic stem cells, coming from a human male.}
\end{enumerate}

The RRBS data for all three cell lines are produced by the Myers Lab at HudsonAlpha Institute for Biotechnology (GEO: GSE27584). The data are already pre-processed and aligned to the \textit{hg19} human reference genome, and can be downloaded from the web accessible database at UCSC. For our analysis, we use the resulting BED files and we ignore strand information. To obtain more accurate methylation level estimates, we pool together all available replicates. To investigate the correlation between DNA methylation profiles and  transcript abundance, we use the corresponding paired-end RNA-seq data produced by Caltech (GEO: GSE33480). The RNA-seq data are pre-processed and mapped to the \textit{hg19} human reference genome using TopHat and transcription quantification, in FPKM (Fragments Per Kilobase transcript per Million mapped reads), is produced using Cufflinks \citep{Trapnell2012}. The RNA-seq data are filtered in order to keep only protein-coding genes. 

To define promoter regions, we extract the TSS from the corresponding RNA-seq data, which are annotated based on both versions v3c and v4 of GENCODE GRCh37. Then, we consider N base pairs upstream and downstream from each TSS, resulting in promoter regions of length 2N base pairs. Since the cell lines are coming from different genders, the sex chromosomes are discarded from further analysis.

\section{Results}

\subsection{Methylation profiles are highly correlated with gene expression}
Initially, we examine whether gene expression levels might be predictable from DNA methylation patterns alone. We therefore extract higher-order features from promoter regions of $\pm 7kb$ around the TSS by learning the corresponding methylation profiles using the BPR observation model. To ensure that the promoter-proximal regions will have enough data to learn reasonable methylation profiles, we discard regions with less than $15$ CpGs, and restrict our attention to regions which exhibit spatial variability in methylation levels. We applied the same pre-processing steps for the three ENCODE cell lines, which resulted in $7093$ promoters for K562, $6022$ for GM12878 and $5753$ for H1-hESC cell line.

We model the methylation profiles using nine RBFs, which results in ten extracted features including the bias term. In addition to these features, we use the goodness of fit in RMSE and the CpG density across each region. We then train the SVM model on the resulting 12 features using a random subset of $70\%$ of the promoter-proximal regions. We test the model's ability to quantitatively predict expression levels on the remaining $30\%$ of the data. Our results show a striking improvement in prediction accuracy when compared to using the mean methylation level as input feature.

\begin{figure}[!tpb]
\centerline{\includegraphics[width=0.78\textwidth]{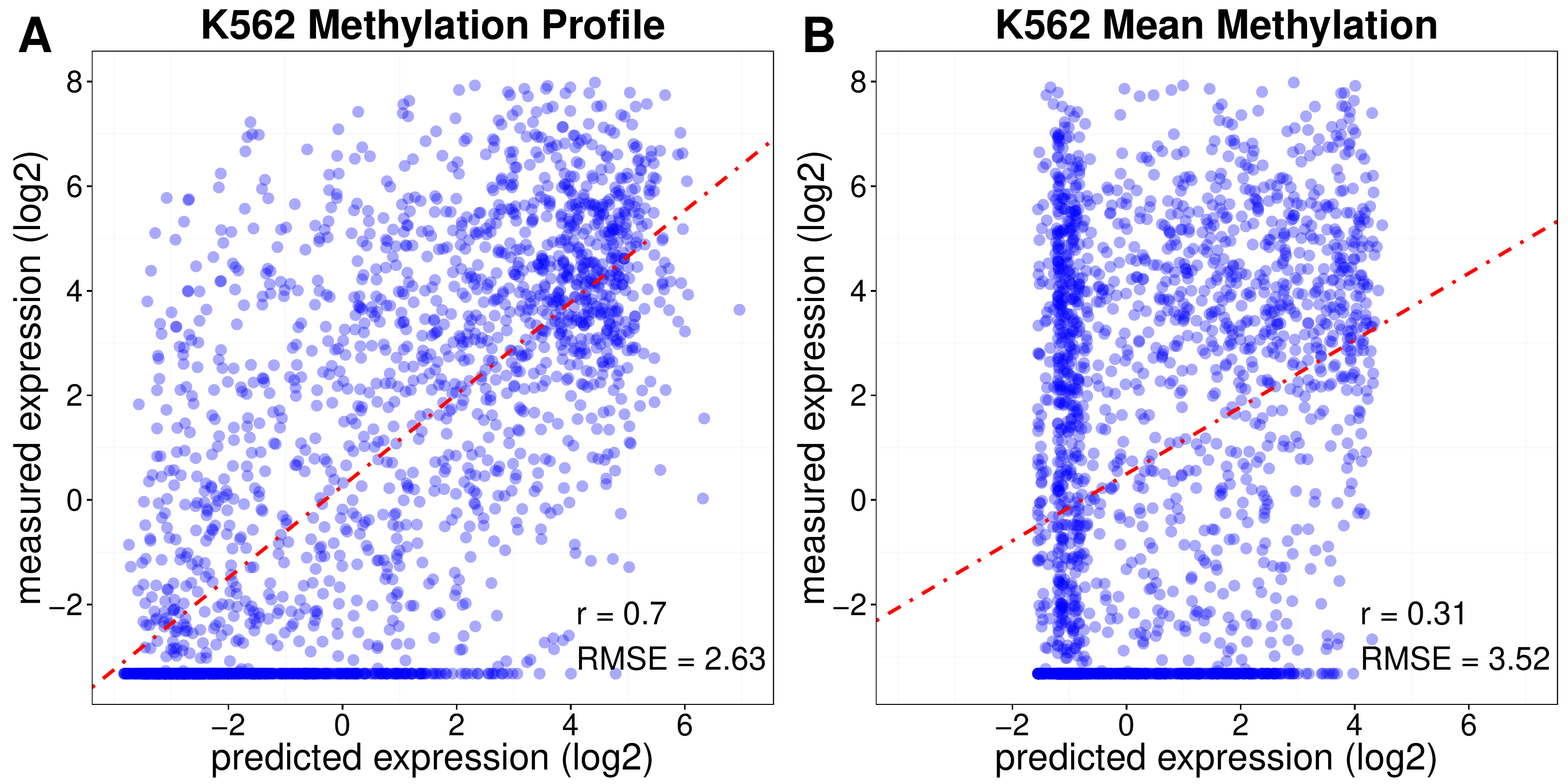}}
\caption{\small{Quantitative relationship between DNA methylation patterns and expression. (\textbf{A}) Scatter plot of predicted gene expression using the BPR model on the x-axis versus the measured (log-transformed) gene expression values for the K562 cell line on the y-axis. Each methylation profile is modelled using nine RBFs. In addition to these features, the SVM regression model uses as input the goodness of fit in RMSE and the CpG density. Each shaded blue dot represents a different gene and the darker the colour, the higher the density of points. The red dashed line indicates the linear fit between the predicted and measured expression values, which are highly correlated (Pearson's $r = 0.7$, p-value $<$ 2.2e-16), indicating a quantitative relationship between methylation profiles across promoter-proximal regions and transcript abundance. The model performance is also assessed by RMSE, which is 2.63. (\textbf{B}) Scatter plot of predicted and measured gene expression values when using the average methylation level as input feature in the SVM model; correlation has decreased substantially ($r$ = 0.31 and RMSE = 3.52).}}
\label{fig:k562-scatter}
\end{figure}

Figure~\ref{fig:k562-scatter}A shows a scatter plot of the predicted and measured expression values for the K562 cell line, with Pearson's $r$ = 0.7 (p-value of t-test $<$ 2.2e-16) and RMSE = 2.63, demonstrating that the shape of methylation patterns across promoter-proximal regions is well correlated to mRNA abundance. Figure~\ref{fig:k562-scatter}B shows the performance of the regression model when using the mean methylation level as input feature. It is evident that this approach cannot capture the diverse patterns present across the promoter regions, leading to poor prediction accuracy ($r$ = 0.31 and RMSE = 3.52). Notice that the mean methylation approach erroneously predicts gene expression values only in the (-2, 4) interval, whereas the BPR model captures more accurately the dynamic range of expression. Interestingly, the mean approach erroneously predicts the majority of genes to have expression value around -1, clearly indicating that summarising DNA methylation by a single average is insufficient to capture the complex relationship with expression. Finally, one should observe the horizontal stripe around -3 on both figures: these are genes whose lack of expression cannot be attributed to DNA methylation patterns, possibly implicating other regulating mechanisms (e.g. histone marks, binding of transcription factors, etc.), or difficulties in the measurement process of RNA-seq experiments (e.g. due to genes having relatively non-unique transcript sequences or multiple promoters).

\begin{figure}[!tpb]
\centerline{\includegraphics[width=0.79\textwidth]{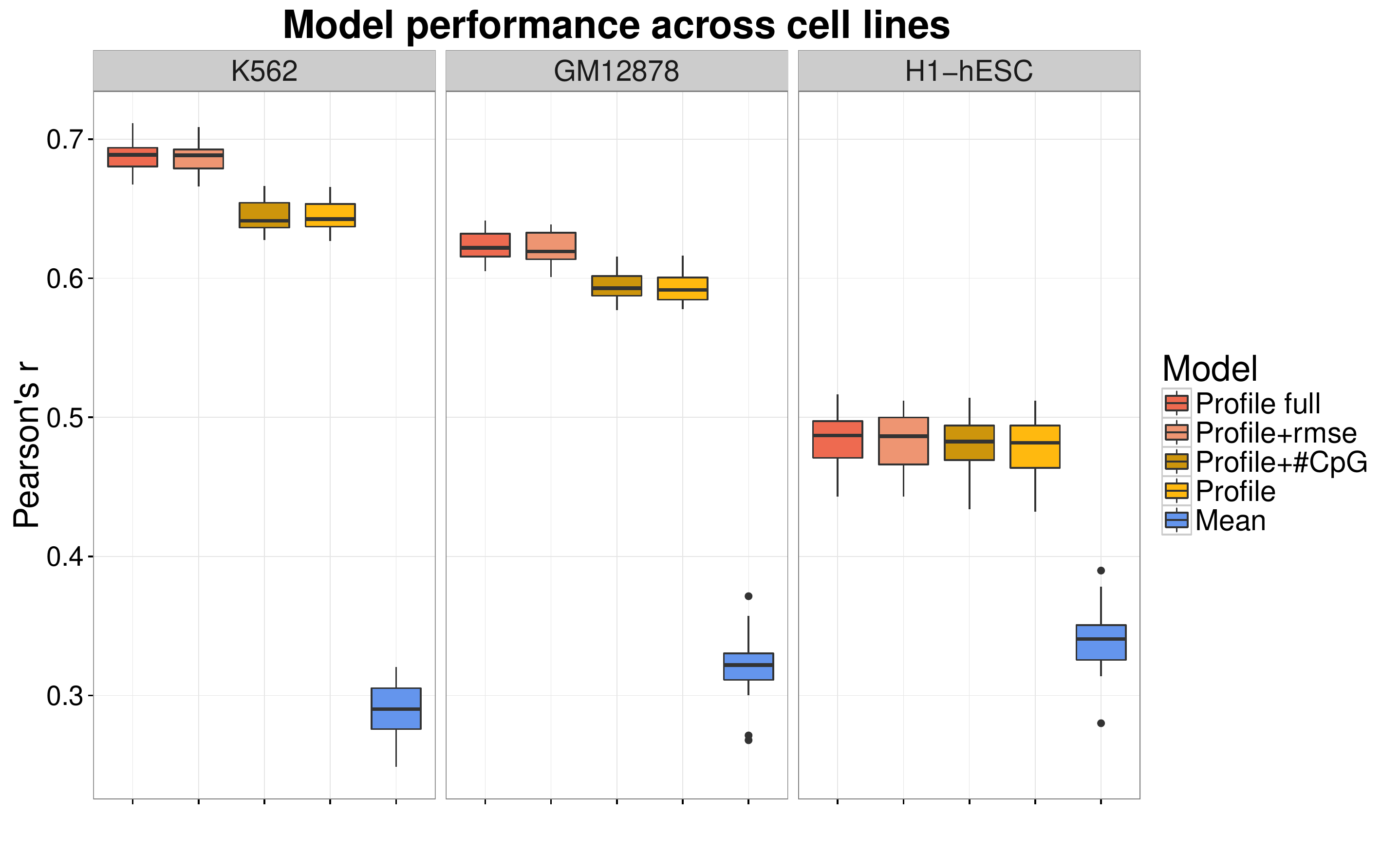}}
\caption{\small{Boxplot of correlation coefficients for the three ENCODE cell lines (K562, GM12878 and H1-hESC) with different input features for the SVM regression. The 'Profile full' model corresponds to the extracted BPR features plus the two additional features. Each boxplot indicates the performance using 20 random splits of the data in training and test sets. Paired Wilcoxon test shows that the high quantitative relationship between the shape of DNA methylation and expression exists in various cell lines, and is significantly better predictor than using the average methylation level (p-value = 8.4e-12). Regarding the two additional features, we observe that the goodness of fit measured in RMSE has a positive impact in correlation, whereas the CpG density does not improve the prediction performance. Paired Wilcoxon tests between K562 and other cell lines, show that K562 has significantly higher prediction accuracy (p-value = 4.8e-05 for both GM12878 and H1-hESC).}}
\label{fig:model-performance}
\end{figure}

We then consider the relative importance of the various features in predicting gene expression: in particular, we are interested in determining whether including goodness of fit or CpG density as covariates has any impact on predictive performance. For each cell line, we learn five SVM regression models, each having a different number of input features. The first four models consider as input the extracted higher-order methylation features with a combination of the two additional features we described in the previous section, whereas the last model takes the average methylation level as input feature. To statistically assess our results, we perform 20 random splits in training and test sets and evaluate the model performance on the corresponding test sets. Figure~\ref{fig:model-performance} shows boxplots of correlation coefficients for the three ENCODE cell lines, where each boxplot indicates the performance of the prediction model on the 20 random splits of the data.  The results demonstrate that by considering higher-order features we can build powerful predictive models of gene expression; and in the case of K562 and GM12878 we have more than 2-fold increase in correlation.
\begin{table}[!b]
\begin{center}
 {\begin{tabular}{@{}c|cccccccc@{}}\toprule
   Cell Line & {$\pm2kb$} & {$\pm3kb$} & {$\pm4kb$} & {$\pm5kb$} & {$\pm6kb$} & {$\pm7kb$} & {$\pm8kb$} & {$\pm9kb$}\\\midrule
   K562 & 0.63 & 0.69 & 0.69 & 0.67 & 0.67 & \textbf{0.70} & 0.67 & 0.67 \\
   GM12878 & 0.62 & 0.62 & 0.64 & 0.61 & 0.62 & \textbf{0.61} & 0.61 & 0.61 \\
   H1-hESC & 0.46 & 0.49 & 0.48 & 0.43 & 0.49 & \textbf{0.50} & 0.47 & 0.49 \\\bottomrule
   \hline
\end{tabular}}
\caption{Pearson's $r$ when considering different promoter region windows. \small{For various length promoter-proximal regions, we show the performance (in Pearson's $r$) of methylation profiles in accurately predicting gene expression. The BPR model has high correlation across all different-length regions for all cell lines considered in this study.}}
\label{Tab:01}
\end{center}
\end{table}

Concentrating on the importance of the additional features for the prediction process, we observe that the addition of CpG density does not have a significant prediction improvement compared to using only the shape of methylation profiles as input features (paired Wilcoxon test p-value = 0.22, 0.18 and 0.02 for K562, GM12878 and H1-hESC, respectively). On the other hand, the goodness of fit of the methylation profile in RMSE has a positive impact on the prediction performance (paired Wilcoxon test p-value = 4.8e-05, 4.8e-05 and 0.0001 for K562, GM12878 and H1-hESC, respectively). Finally, we explore the importance of considering different promoter region windows. Table~\ref{Tab:01} shows Pearson's $r$ when considering various length promoter-proximal regions around the TSS. In general, the BPR model maintains its high predictive power across all cell lines for all different-length regions.

\subsection{Methylation profiles are predictive of gene expression across different ENCODE cell lines}
\begin{figure}[!tp]
\centerline{\includegraphics[width=0.75\textwidth]{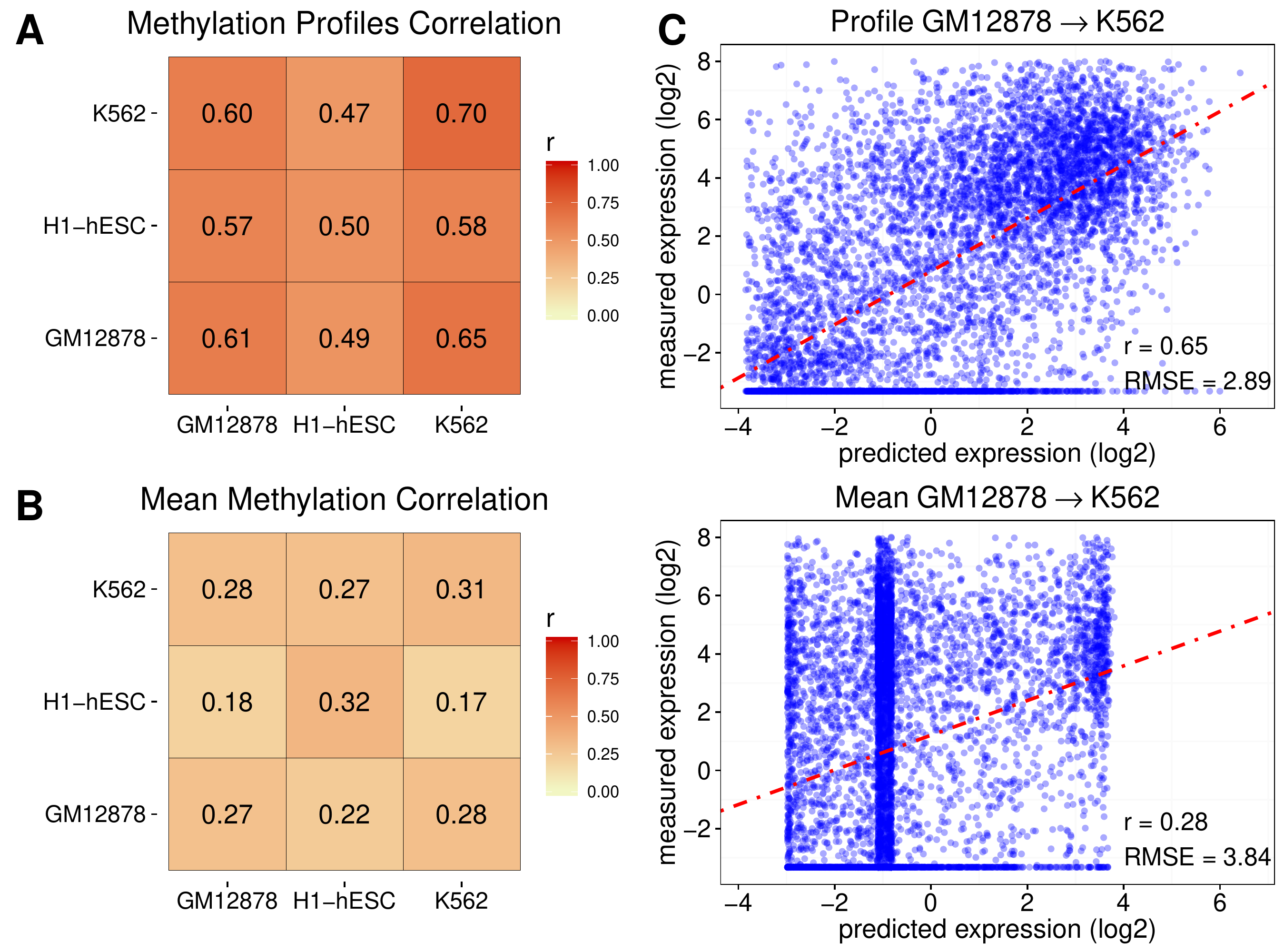}}
\caption{\small{Prediction accuracy across different cell lines. (\textbf{A}) Confusion matrix of correlation coefficients across cell-lines when using the BPR model with nine RBFs as input features to the regression model. Each $(i,j)$ entry of the confusion matrix, corresponds to training a regression model from $i^{th}$ cell line and predicting gene expression levels for the $j^{th}$ cell line. The colour of the confusion matrix corresponds to Pearson's $r$ value, the darker the colour the higher the correlation. (\textbf{B}) The corresponding correlation coefficients when using the mean methylation level as input feature to the regression model. Comparing both confusion matrices, it is evident that the methylation profile approach is more powerful in predicting expression levels across different cell lines. (\textbf{C}) Application of the model learned from GM12878 cell line to predict expression levels of the K562 cell line, using methylation profiles (top) and mean methylation levels (bottom) as input features.}}\label{fig:cross-cell-line}
\end{figure}

We showed that gene expression is effectively predicted from the BPR model by using higher-order methylation features among various cell lines. Next, we further explore if the proposed model maintains predictive power across different cell lines. That is, we apply the regression model trained on one cell line to predict expression levels in another cell line, by using the learned methylation profiles in those cell lines as input features to the regression model. Figures~\ref{fig:cross-cell-line}A-B show confusion matrices of correlation coefficients for the cross-cell line prediction process, using the BPR model and the mean methylation level approach, respectively. Figure~\ref{fig:cross-cell-line}C shows an example of applying the model learned from GM12878 methylation patterns to predict expression levels of the K562 cell line. The BPR model effectively predicts gene expression ($r$ = 065 and 0.49 for predicting K562 and H1-hESC, respectively), while, the mean methylation approach provides a poor estimate of correlation ($r$ = 0.28 and 0.22 for predicting K562 and H1-hESC, respectively).

The results indicate that the quantitative relationship between DNA methylation profiles and mRNA abundance is not cell line specific, but that the model captures patterns of association between methylation and expression which hold across different cell lines. Although the proposed models have high prediction accuracy across all cell lines, the H1-hESC cell line shows consistently weaker correlations. This finding is in line with recent studies, reporting weaker correlations of gene expression and chromatin features for the H1-hESC cell line \citep{Dong2012}, and with observations that mRNA-encoding genes in stem cells are transcriptionally paused during cell differentiation \citep{Min2011}.

\subsection{Clustering DNA methylation profiles across promoter - proximal regions}
\begin{figure*}[!tpb]
\centerline{\includegraphics[width=0.99\textwidth]{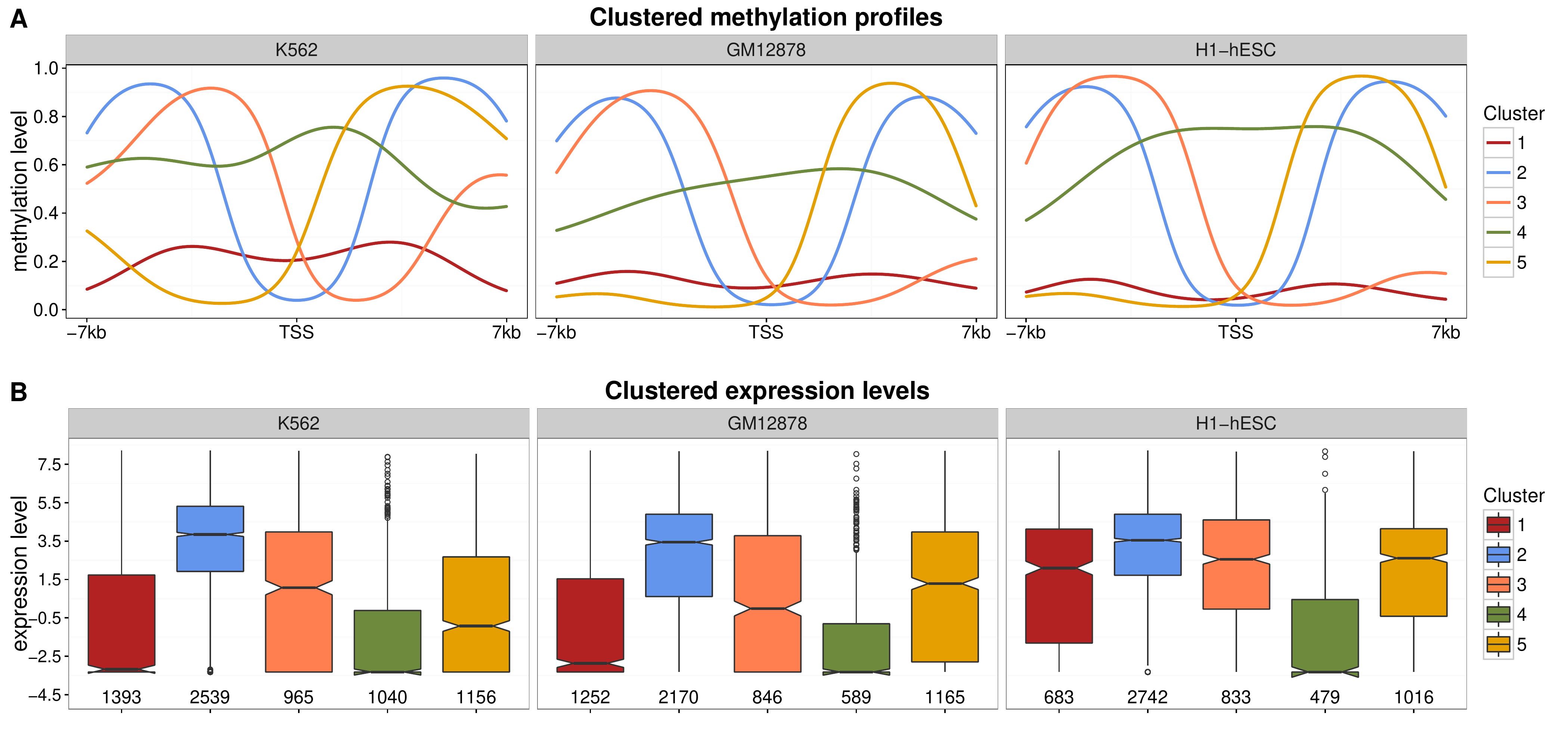}}
\caption{\small{Clustering DNA methylation profiles across promoter-proximal regions. (\textbf{A}) Five clustered methylation profiles over $\pm 7$kb promoter region w.r.t. TSS in the direction of transcription for the three ENCODE cell lines (K562, GM12878 and H1-hESC). Each methylation profile is modelled using four RBFs. Comparing the clustered profiles it is evident that there are five prototypical methylation shapes across the cell lines. (\textbf{B}) Boxplots with the corresponding expression levels of the protein-coding genes assigned to each cluster for each of the three cell lines. The colours match with the clustered methylation profiles shown above. The numbers below each boxplot correspond to the total number of genes asssigned to each cluster. T-test shows that the U-shape methylation profiles (Cluster 2, blue) correspond to significantly higher expression (p-value $<$ 2.2e-16) compared to the expression of genes assigned to the remaining methylation profiles.}} \label{fig:clusters}
\end{figure*}

We next use the higher order methylation features to cluster DNA methylation patterns across promoter-proximal regions and examine whether distinct methylation patterns across different cell lines are associated to gene expression levels. We apply the same pre-processing steps described in the previous sections and we consider genomic regions of $\pm 7$kb around the TSS. We use the Bayesian Information Criterion (BIC) to set the number of clusters to five. We model the methylation profiles at a slightly lower spatial resolution, using four RBFs, as we are interested in capturing broader similarities between profiles, rather than fine details. Figure~\ref{fig:clusters}A shows the five distinct methylation profiles that were learned from each cell line after applying the EM algorithm. To investigate the association of promoter methylation profiles and transcription, in Figure~\ref{fig:clusters}B we show boxplots with the corresponding mRNA expression values that are assigned to each cluster for each cell line. From the resulting methylation profile clusters, we seek to characterize the common features that are responsible for the corresponding mRNA abundance. 

As expected, clusters corresponding to hyper-methylated regions (Cluster 4, green) are associated with repressed genes across all cell lines, confirming the known relationship of DNA methylation around TSSs with gene repression. Also, two distinct patterns emerge: an S-shape profile (Cluster 5, yellow) with hypo-methylated CpGs upstream of TSS, which become gradually methylated at the gene body, and the reverse S-shape pattern (Cluster 3, orange). Genes associated with these profiles have intermediate expression levels for K562 and GM12878, and relatively high expression for H1-hESC. The most interesting pattern is the U-shape methylation profile (Cluster 2, blue), with a hypo-methylated region around the TSS surrounded by hyper-methylated domains. These profiles are associated with high transcriptional activity at their associated genes across all cell lines (t-test p-value $<$ 2.2e-16 for all paired cluster comparisons across cell lines).  Surprisingly, uniformly low-methylated domains (Cluster 1, red) seem in general to be repressed, except from the H1-hESC cell line, suggesting a different type of relationship between DNA methylation and expression in embryonic stem cells. The clustering analysis confirms, in a complementary way, that DNA methylation profiles and transcriptional process are tightly connected to each other, and this relationship can be generalized across all cell lines considered in this study.

To provide a biological insight on the potential methylation mechanisms that regulate transcription, we consider the purity of the clustering across different cell lines, i.e., which fraction of genes assigned to a certain cluster in a certain cell line are assigned to the same cluster in the other cell lines. Surprisingly, around 68\% of the genes assigned to the U-shape profile are present in all three cell lines, while the intersection of genes assigned to the other clusters ranges between 20\% to 40\%.  Interestingly, the promoter-proximal regions clustered to the U-shape methylation profile are dominated by CGIs. Of all common promoters assigned to the U-shape profiles, 95.6\% are CGI associated. Not surprisingly, hyper-methylated promoters are only 35.7\% CGI associated, however uniformly low-methylated promoters are 65.9\% CGI associated. This suggests, that promoters associated with totally unmethylated CGIs surrounded by hyper-methylated domains are transcriptionally active across cell lines. Indeed, we find that 35\% of the U-shape profile genes are associated with a curated set of housekeeping genes \citep{Eisenberg2013}. On the contrary, only a small fraction of genes assigned to hyper-methylated domains or uniformly low-methylated domains are housekeeping genes (1.4\% and 17.7\% respectively). Finally, around 22\% of the genes assigned to the S-shape and reverse S-shape profiles are associated with housekeeping genes. 

\section{Discussion}
Alterations in DNA methylation are associated with regulatory roles and are involved in many diseases, most notably cancer \citep{Baylin2011}. Therefore, unravelling the function of DNA methylation and its relationship to transcription, is essential for understanding biological processes and for developing biomarkers for disease diagnostics \citep{Laird2003}.

Our results demonstrate that representing methylation patterns by their average level is insufficient to understand the link between DNA methylation and expression, and one should consider the shape of the methylation profiles at the vicinity of the promoters. The contributions of this paper are twofold. First, we introduced a generic modelling approach to quantitate spatially distributed methylation profiles via the BPR model. The BPR features enabled us to build a powerful predictive model for gene expression in various cell lines which more than doubled the predictive accuracy of current methods based on average methylation levels. 

Second, we have shown how the BPR features can be used in downstream analyses by clustering spatially similar methylation profiles. We revealed five distinct groups of methylation patterns across promoter regions that are well correlated with gene expression and are well reproducible across different cell lines. Some of these patterns recapitulate existing biological knowledge. The U-shape methylation profile, consisting of hypo-methylated CGIs followed by hyper-methylated CGI shores, has been identified in different studies, and is termed as 'canyon' \citep{Jeong2014} or 'ravine' \citep{Edgar2014}. Our findings are in line with \cite{Edgar2014}, where ravines are in general positively correlated with mRNA abundance. Since, the main difference of the U-shape methylation profile and the uniformly low-methylated profile is the CGI shore methylation, our results support the hypothesis that hyper-methylation on the edges of CGIs enhances transcriptional activity.

The existence of U-shape methylation profiles may help to explain observations that the methylation of gene body was sometimes positively correlated with transcript abundance \citep{Varley2013, Lou2014}. We hypothesize that these regions may correspond to U-shape methylation profiles, or a mixture of U-shape and S-shape methylation profiles. Another relevant study, showed that hyper-methylation of CGI shores on the mouse genome was associated with increased DNMT3A activity, which resulted in positive correlation with transcriptional activity; indicating that methylation outside of CGIs may by used for maintaining active chromatin states for specific genes \citep{Wu2010}. 

In this study, we focused on RRBS data, however, given the considerable robustness of the BPR model to low coverage, we expect that it may also be well suited for Whole Genome Bisulphite Sequencing data, which have the advantage of providing a more comprehensive coverage of CpG sites genome-wide. As an extension of this analysis, further work could include building a model to relate differential methylation profiles with differential gene expression levels, and evaluate the importance of profile changes in regulation of gene expression across different cell types. More generally, it is increasingly clear that transcriptional activity is regulated by a complex and still incompletely understood interaction network of molecular players, including DNA methylation, histone modifications and transcription factor binding. Several recent computational studies have highlighted the dependencies between these players \citep{Dong2012, Benveniste2014}. The BPR model provides an effective way of recapitulating DNA methylation patterns using higher order features, and may therefore play an important role in building more effective integrative models of high-throughput data.


\section*{Acknowledgements}
We thank Duncan Sproul for valuable comments and discussion. \emph{Funding}: CAK is supported in part by the EPSRC Centre for Doctoral Training in Data Science, funded by the UK Engineering and Physical Sciences Research Council (grant EP/L016427/1) and the University of Edinburgh. GS is funded by the European Research Council through grant MLCS306999.


\bibliographystyle{natbib}
\bibliography{ECCB2016}

\end{document}